\documentclass{aa}
\usepackage{graphicx}
\usepackage{txfonts}
\begin{document}
   \title{The X-ray spectrum of the black hole candidate GX339-4 in a low state}

   \author{A.Corongiu
          \inst{1}\fnmsep\inst{2}\fnmsep\inst{3}
          \and
          L.Chiappetti\inst{1}
          \and
          F.Haardt\inst{4}
          \and
          A.Treves\inst{4}
          \and
          M.Colpi\inst{5}
          \and
          T.Belloni\inst{6}
          }

   \offprints{A.Corongiu {\email {corongiu@ca.astro.it}}}

   \institute{ IASF-CNR,
               Sezione di Milano ``G.Occhialini'',
               via Bassini 15, I-20133 Milano
         \and
              Dipartimento di Scienze Fisiche, Universit{\`a} degli Studi di Milano,
              via Celoria 16, I-20133 Milano
         \and
             INAF - Osservatorio Astronomico di Cagliari,
             Loc. Poggio dei Pini, strada 54, I-09012 Capoterra
         \and
             Dipartimento di Scienze, Universit{\`a} dell'Insubria/Como,
             via Valleggio 11, I-22100 Como
         \and
             Dipartimento di Fisica, Universit{\`a} di Milano Bicocca,
             piazza delle Scienze 3, I-20126 Milano
         \and
             INAF - Osservatorio Astronomico di Brera,
             via E. Bianchi 46, I-23807 Merate
             }

   \date{Received April 22, 2003 / accepted June 12, 2003}

   \abstract{ We report on four observations of the
black hole candidate GX~339-4 collected in the spectral range
0.4-140 keV with the BeppoSAX satellite during 1997.
All our spectra are typical of a black hole candidate in
its low-hard state, i.e. are dominated by the hard tail of
the emission from the corona. Although the direct
emission from the accretion disk in the soft X-ray range is weak,
spectra are well described only with models that take
into consideration the Compton reflection of hard X-rays
from the corona onto the surface of an accretion disk
of non negligible extent.
   }

   \maketitle

\section{Introduction}

GX~339-4, discovered in 1971 by Markert et
al. (\cite{marka}, \cite{markb}), was
proposed to be a black hole candidate by Samimi
et al. (\cite{samimi}), because of its spectral
features and its short term variability
similar to sources like Cyg X-1.

\begin{figure*}
   \centering
   \includegraphics[angle=90,width=17cm]{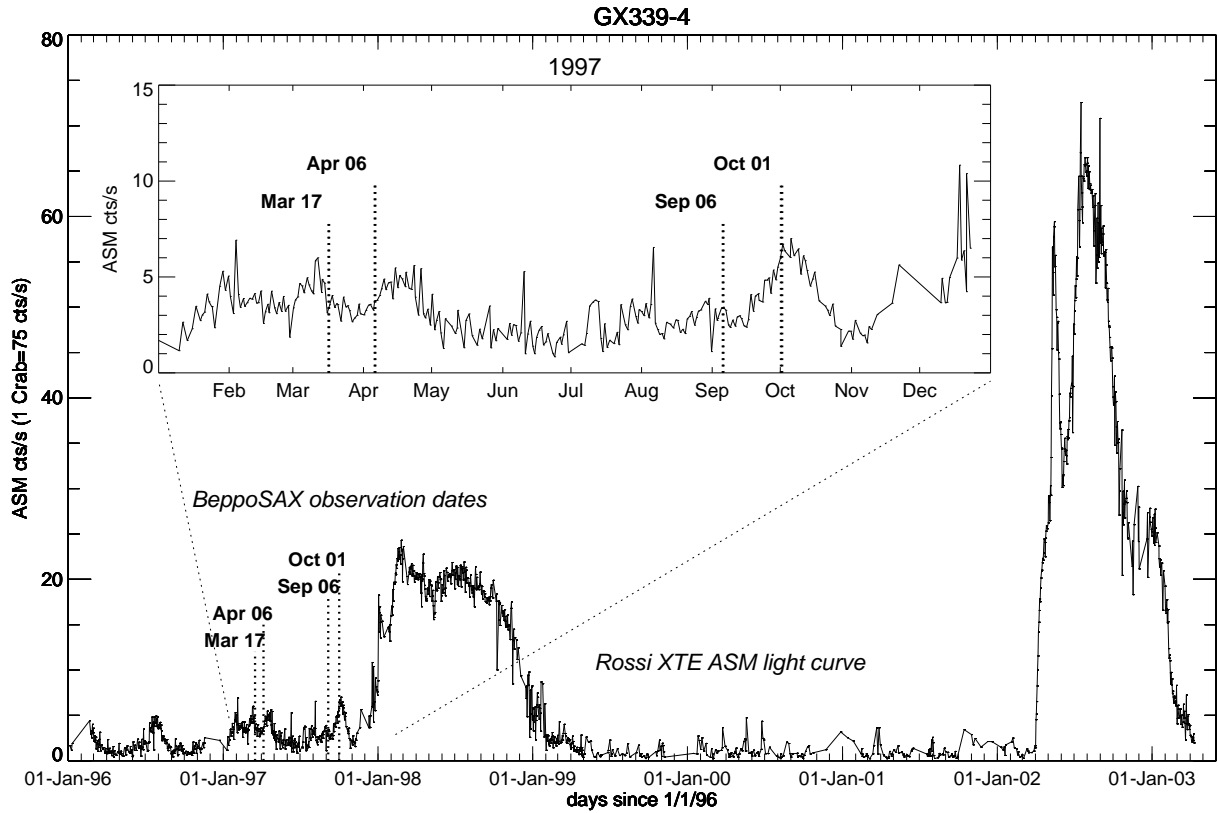}
   \caption{Historical GX~339-4 light curve as observed by the All Sky Monitor
            onboard Rossi XTE. The dashed vertical lines in the main panel and
            in the inset indicate the epoch of our BeppoSAX
            observations. }
    \label{FigXTE}%
    \end{figure*}

GX~339-4 was observed in five different
spectral states, each one with well defined features
(Tanaka \& Lewin \cite{tanaka}).
Most observations detected the source in the hard-low
state, and often with such a low count rate to properly call it
an off state; less often the source was detected in the
high-soft state, only twice in the very high
state (before 2001) and twice in the intermediate
state (for a recent re-consideration of the nature of the
intermediate state see Homan et al. \cite{homan}).
In the first half of 1999, the source entered an extended off-state,
and it remained undetectable by the All-Sky Monitor on board
RossiXTE (see Kong et al. \cite{kong}, Corbel et al. \cite{Corbel}).
During 2002 GX~339-4 started a new outburst (Smith et al. \cite{smith}) with
more state transitions (Belloni et al. \cite{belloni2})
The light curve from the RXTE/ASM is shown in
Fig.~\ref{FigXTE}.

In its low state GX~339-4 shows 
power law spectra, with photon index $\Gamma$=1.62-2.30
(Dolan et al. \cite{dolan}; Bouchet et al. \cite{bouchet}; Ueda et
al. \cite{ueda}; Mendez \& van der Klis \cite{mendez}; Zdziarski et
al. \cite{zdziarski}; Belloni et al. \cite{belloni}; Wilms et al. \cite{wilms}).
Higher values for $\Gamma$ were detected in
spectra observed up to 200 keV ($\Gamma=2.0\pm0.3$;
Dolan et al. \cite{dolan}) and to 1.3 MeV (Bouchet et al.
\cite{bouchet}). Most observations in the low state
indicate the presence of the fluorescence Fe line
(Ueda et al. \cite{ueda}; Zdziarski et al. \cite{zdziarski}; Dolan et
al. \cite{dolan}), with an equivalent width of about 50-100 eV.

Several authors (Zdziarski et al. \cite{zdziarski};
Wilms et al. \cite{wilms}; Revnivtsev et al. \cite{revnivtsev};
Nowak et al. \cite{nowak}; Wardzi\'nski et al. \cite{wardzinski})
give indication of a reflection component in the X spectra,
described according to the model of Magdziarz \&
Zdziarski (\cite{magdziarz}) and provide evidence for the Fe
abundance being greater by a factor 2-3 relative to solar
abundance.

GX~339-4 high states have also well defined spectral
features. In this state the spectrum shows a soft
component, due to the thermal emission by an accretion
disk (Makishima et al. \cite{makishima}; Belloni et al.
\cite{belloni}), described by a multicolour blackbody model
(Makishima et al. \cite{makishima}). Inner temperatures of the
disk were estimated to be 0.77 keV (Makishima et al.
\cite{makishima}) and 0.63-0.72 keV (Belloni et al. \cite{belloni}). The
high energy part of the observed spectra has a powerlaw
behaviour, with a photon index $\Gamma$ of 2.0
(Makishima et al. \cite{makishima}) and of 2.12-2.57 (Belloni et
al. \cite{belloni}).

In the high state the fluorescence Fe line
at 6.4 keV is present, with an equivalent width of 319-350 eV.
There is also evidence for an absorption edge,
although its central energy is uncertain:
Makishima et al. (\cite{makishima}) report
a value of 8.8 keV, while Belloni et al.
(\cite{belloni}) one of 6.69 keV.

The source is very active also in the radio band, with radio
flux strictly correlated to the X-ray intensity (Corbel et al. \cite{Corbel};
Markoff et al. \cite{Markoff}).

In this paper we report results from four 
observations of GX~339-4, taken during 1997 with the
BeppoSAX satellite, when the source was in a low-hard
state. Using all four narrow field instruments available
in the satellite scientific payload, we obtained and analysed
data in the 0.4--140.0 keV band. In Sect. 2 we present the four
observations with a summary of the reduction procedure;
in Sect. 3 we describe the data analysis and in Sect. 4 we discuss
our results.

   \begin{table*}
      \caption[]{Journal of BeppoSAX observations}
         \label{TabJou}
         \begin{tabular}{llllllllllllll}
            \hline
            \hline
            \noalign{\smallskip}
            Number & Start~epoch & LECS & LECS & MECS3 & MECS3 & HPGSPC & HPGSPC & PDS & PDS \\
            & (UT) & exposure & count rate & exposure & count rate & exposure & count rate & exposure & count rate\\
            \noalign{\smallskip}
            \hline
            \noalign{\smallskip}
            1 & 17~Mar~07:08 & 4888 s & 6.22$\pm$0.04 & 10108 s & 6.92$\pm$0.03 & 4262 s & 21.5$\pm$0.3 & 4407+4273 s & 18.6$\pm$0.1 \\
            2 & 06~Apr~10:13 & 7302 s & 5.83$\pm$0.03 & 12286 s & 6.10$\pm$0.02 & 5628 s & 18.8$\pm$0.2 & 5631+5626 s & 16.8$\pm$0.1 \\
            3 & 06~Sep~05:51 & 5947 s & 3.96$\pm$0.03 & 16858 s & 4.72$\pm$0.02 & 7432 s & 15.7$\pm$0.2 & 7976+7906 s & 13.9$\pm$0.1 \\
            4 & 01~Oct~17:36 & 5079 s & 9.01$\pm$0.04 & 25046 s & 9.81$\pm$0.02 & 11006 s & 27.6$\pm$0.1 & 10897+11018 s & 22.0$\pm$0.1 \\
            \noalign{\smallskip}
            \hline
         \end{tabular}
\begin{list}{}{}
\item[a] Count rates measured in the energy bands used for the fits, namely: 
0.4-4.0 keV (LECS); 4.0-10.0 keV (MECS1); 1.8-10.0 keV (MECS2 and MECS3); 
7.0--70.0 keV (HPGSPC); 20.0--140.0 keV (PDS).
\item[b] For MECS we indicate in the table the count rate for unit MECS3.
Unit MECS2 has a similar count rate, while unit MECS1 (where present, i.e. in obs 1 \& 2)
has a slightly lower count rate due to thicker ion shield.
\item[c] For the PDS instrument we quote the exposure time as the sum of the two instrument halves
while the count rate is the net count rate in the 4 units of which 2 look at the source at any one time
for the quoted exposure times
\end{list}
   \end{table*}


\section{Observations and data reduction}

We observed GX~339-4 with BeppoSAX (Boella et al. \cite{boella} and references therein)
at four epochs during 1997, with the primary aim to
study the spectral variability on the 1-month and 6-month time scale.
A journal with basic data is reported in Table~\ref{TabJou}, while the
state of the source at the epoch of our observations with respect to the
historical light curve is visible in Fig.~\ref{FigXTE}.
Preliminary results on the first two observations have been previously
reported by Chiappetti et al. (\cite{lucio0}).

From the figure it is apparent that we observed the source at only slightly
different intensity levels during a low state. 

We detected GX~339-4 with all the four Narrow Field Instruments on board
BeppoSAX, obtaining a simultaneous coverage with an excellent signal to noise 
from 0.4 to 140 keV.
MECS (Medium Energy Concentrator Spectrometer) observations used
all three MECS units for the first two observations, and only two
units for the last two, due to a failure of unit MECS1 in late spring 1997.

Data reduction for MECS, HPGSPC (High Pressure Gas Scintillation
Proportional Counter) and PDS (Phoswich Detector System) was performed using 
the XAS software (Chiappetti \& Dal Fiume \cite{lucio1}),
while for LECS (Low Energy Concentrator Spectrometer) data we used the 
cleaned event files supplied by the ASI Science Data Centre (ASDC). In general
the reduction followed the standard prescriptions described e.g. in
Chiappetti et al. (\cite{lucio2}), with only slight differences and peculiarities
due to the higher intensity of GX~339-4. A summary is given below.
The reader not interested in reduction details may wish to proceed directly
to Sect. 3.

We examined the light curves of GX~339-4 during our (relatively short) observations,
but, since we found no indication of large variability, we proceeded accumulating
a single spectrum per instrument for each observation.

LECS gross spectra were extracted from event files
in a circular area of radius 8 arcmin. Background
subtraction was performed linking each data file to
a standard background file (the latest available at
ASDC\footnote{calibration files repository:
{\tt ftp.asdc.asi.it/pub/sax/cal/}}
in November 1999). ASDC standard response matrices
were also used. In our analysis we used data in the
0.4--4.0 keV energy band.

A larger extraction radius of 13 arcmin was used for MECS.
Background subtraction was performed using a 1120 ks background file
derived from blank field pointings taken during the satellite
Performance Verification phase.
Response matrices were generated with XAS according to the source
location in the field of view.
In our analysis we used data in the 4.0--10.0 keV energy band for MECS1, 
in the 1.8--10.0 keV
energy band for MECS2 and MECS3.

HPGSPC background subtraction was performed
using data taken during our observations (but not simultaneously) with
the rocking collimator offset in the negative direction.
ASDC standard response matrices were used.
In our analysis we used data in the 7.0--70.0 keV energy band.

PDS background subtraction used simultaneous data taken
with the rocking collimators in offset position (Frontera et al. \cite{frontera}). 
In consideration of the source strength the threshold for spurious spike
filtering has been 32 cts/s (instead of the customary 25 cts/s), and the PSA
correction was not applied.
Response matrices were computed with XAS
from the configuration recorded in the spectra.
In our analysis we used data in the 20.0--140.0 keV energy band.

Given the brightness of the source, the systematics in the calibrations
of the instrument play an important role when matching spectra taken
with different instruments.
This is customarily handled using the standard response matrices and
introducing cross-normalisation constants in the spectral fit.
Unit MECS3 was used as reference (constant fixed to unity).
Obtained cross-normalisations for the other instruments are in good agreement with
values widely reported in the literature and recommended by the BeppoSAX team.


\section{Data analysis}

Spectral fits were performed with the software package XSPEC
v. 10.0. In the remainder of the paper we refer to all additive or
multiplicative spectral components, as well as their parameters,
with the mnemonic used in the XSPEC handbook.

Preliminary fits were made with the model
{\tt WABS~(DISKBB + POWERLAW)}, to account for the
direct emission from the two main structures around a
black hole accreting matter from a companion in a   
binary system: the optically thick, geometrically
thin, accretion disk ({\tt DISKBB}, Makishima et al. \cite{makishima}) and an
optically thin Comptonising corona 
({\tt POWERLAW}). The component
{\tt WABS} describes photoelectric absorption
due to the interstellar matter (Morrison \& McCammon \cite{morrison}, 
Anders \& Ebihara \cite{anders}).

\begin{figure}
   \centering
   \includegraphics[width=8.5cm]{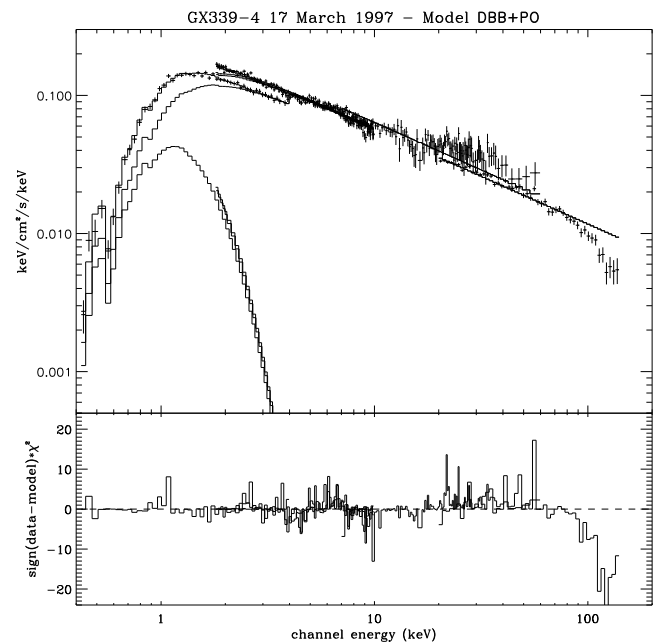}
    \caption{Result of the {\tt WABS (DISKBB + POWERLAW)} fit to the
     spectrum of Mar 17 : unfolded spectrum in top panel,
     contribution to $\chi^2$ in bottom panel.}
    \label{FigWDP}
    \end{figure}

Values of the $\chi^2$ (see Table~\ref{TabWDP}) indicate that such a
simple model does not describe satisfactorily the
observed spectra, and 
Fig.~\ref{FigWDP} (bottom panel) shows that the main
deviations from the model are in the high energy
tail of the spectra: the model underestimates the
observed spectra at energies around 30-60 keV, and
overestimates the spectra at higher energies.

\begin{table*}
      \caption[]{Spectral fits with the
      {\tt WABS (DISKBB + POWERLAW)} model.}
      \label{TabWDP}
\begin{tabular}{lllllll}
\hline
\hline
\noalign{\smallskip}
\multicolumn{1}{l}{Component}&
\multicolumn{1}{l}{Parameter}&
\multicolumn{1}{l}{Unit}&
\multicolumn{1}{l}{17 March}&
\multicolumn{1}{l}{6 April}&
\multicolumn{1}{l}{6 September}&
\multicolumn{1}{l}{1 October}\\
\noalign{\smallskip}
\hline
\noalign{\smallskip}
{\tt WABS} & N$_{\rm{H}}$ & 10$^{21}$ cm$^{-2}$ & 5.00$_{-0.34}^{+0.36}$ & 5.08$_{-0.22}^{+0.28}$ &
4.63$_{-0.30}^{+0.35}$ & 6.04$_{-0.24}^{+0.25}$\\ 
{\tt DISKBB} & T$_{\rm{in}}$ & keV & 0.301$_{-0.022}^{+0.025}$ & 0.351$_{-0.026}^{+0.030}$ &
0.360$_{-0.040}^{+0.051}$ & 0.363$_{-0.018}^{+0.010}$ \\
{\tt DISKBB} & K$_{\rm{DBB}}$ & (km/10 pc)$^{2}$ & 3272.$_{-1489}^{+2466}$ & 1147.$_{-468}^{+768}$ &
501$_{-290}^{+613}$ & 1786.$_{-580.}^{+791.}$\\
{\tt POWERLAW} & $\Gamma$ & & 1.691$_{-0.013}^{+0.013}$ & 1.695$_{-0.012}^{+0.013}$ &
1.666$_{-0.013}^{+0.013}$ & 1.773$_{-0.007}^{+0.007}$\\
{\tt POWERLAW} & K$_{\rm{PO}}$ & keV/cm$^{2}$ s keV (at 1 keV) & 0.286$_{-0.007}^{+0.007}$ & 0.252$_{-0.006}^{+0.006}$ &   
0.188$_{-0.004}^{+0.004}$ & 0.454$_{-0.007}^{+0.006}$\\   
\noalign{\smallskip}
\hline
\noalign{\smallskip}
& $\chi^{2}$ & & 706.78 & 791.97 & 685.33 & 2227.68\\
& $\chi^{2}$/D.o.F. & & 1.78 & 1.97 & 1.94 & 5.64\\
\noalign{\smallskip}  
\hline
\noalign{\smallskip}
\end{tabular}
\begin{list}{}{}
\item[a] Errors quoted are at the 90\% confidence level.
\end{list}
\end{table*}

A first possible interpretation of such deviations would be that
the spectra have a high energy cutoff, which has not
been taken into account in such a simple model : 
the underestimate at medium energies could be
interpreted as the attempt to compensate the lack of
a cutoff in the model. A
less important deviation is seen at energies around
the fluorescence Fe line. 

\begin{figure}
   \centering
   \includegraphics[          width=8.5cm]{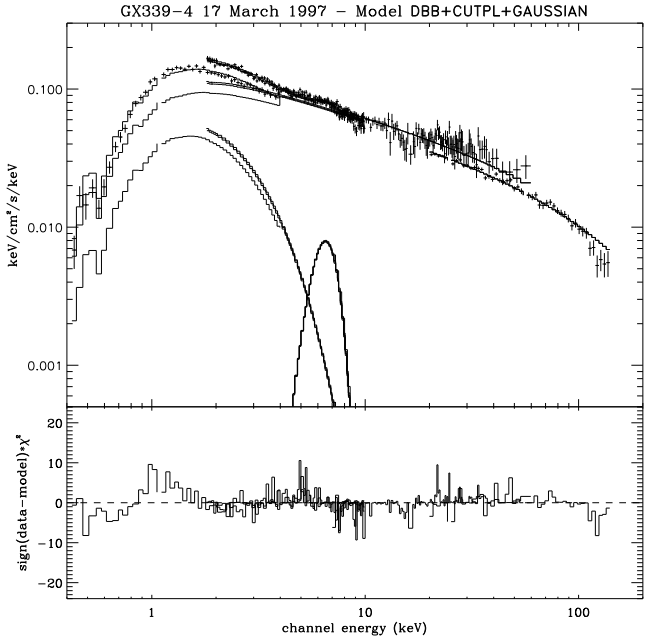}
   \caption{Result of the {\tt WABS (DISKBB + CUTOFFPL + GAUSSIAN)} fit to the
            spectrum of Mar 17 : unfolded spectrum in top panel,
            contribution to $\chi^2$ in bottom panel.}
    \label{FigDCG}
    \end{figure}

Therefore, a new series of fits were
performed with the model {\tt WABS~(DISKBB + CUTOFFPL + GAUSSIAN)}:
the component {\tt CUTOFFPL} was
introduced to account for the high
energy cutoff in the hard tail of the spectra and the
component {\tt GAUSSIAN} for the profile of the Fe
fluorescence line. The Fe line energy was kept
fixed to the value of 6.4 keV (see Table~\ref{TabDCG}). $\chi^2$
values improved for all observations, but they were still
too high to indicate good fits. 
Fig.~\ref{FigDCG} (bottom panel) shows the presence of non negligible
modulations, indicating that this improved model cannot describe
satisfactorily the shape of the observed spectrum. 
Fig.~\ref{FigDCG} (top panel) shows clearly that data remain above the model curve
in the range about 20--50 keV and are definitely below the model in the hard tail.

\begin{table*}
      \caption[]{Spectral fits with the {\tt WABS (DISKBB + CUTOFFPL + GAUSSIAN)} model}
      \label{TabDCG}
\begin{tabular}{lllllll}
\hline
\hline
\noalign{\smallskip}
\multicolumn{1}{l}{Component}&
\multicolumn{1}{l}{Parameter}&
\multicolumn{1}{l}{Unit}&
\multicolumn{1}{l}{17 March}&
\multicolumn{1}{l}{6 April}&
\multicolumn{1}{l}{6 September}&
\multicolumn{1}{l}{1 October}\\
\noalign{\smallskip}
\hline
\noalign{\smallskip}
{\tt WABS} & N$_{\rm{H}}$ & 10$^{21}$ cm$^{-2}$ & 3.18$_{-0.29}^{+0.70}$ & 3.83$_{-0.25}^{+0.26}$ &
2.88$_{-0.17}^{+0.18}$ & 3.73$_{-0.12}^{+0.13}$\\
{\tt DISKBB} & T$_{\rm{in}}$ & keV & 0.757$_{-0.240}^{+0.096}$ & 0.664$_{-0.064}^{+0.075}$ &
0.937$_{-0.047}^{+0.043}$ & 0.877$_{-0.022}^{+0.021}$ \\
{\tt DISKBB} & K$_{\rm{DBB}}$ & (km/10 pc)$^{2}$  & 50.24$_{-13.77}^{+120.44}$ & 81.2$_{-24.6}^{+41.0}$ &
19.4$_{-2.6}^{+3.3}$ & 72.0$_{-5.8}^{+6.6}$\\
{\tt CUTOFFPL} & $\Gamma$ & & 1.446$_{-0.082}^{+0.127}$ & 1.463$_{-0.054}^{+0.045}$ &
1.283$_{-0.053}^{+0.052}$ & 1.262$_{-0.031}^{+0.030}$\\
{\tt CUTOFFPL} & E$_{\rm{cut}} $ & keV & 150.7$_{-31.1}^{+91.2}$  & 147.9$_{-22.5}^{+27.3}$ &
105.5$_{-11.9}^{+14.9}$ & 74.5$_{-3.8}^{+4.2}$\\
{\tt CUTOFFPL} & K$_{\rm{cutpl}}$ & keV/cm$^{2}$ s keV (at 1 keV) & 1.77$_{-0.30}^{+0.56}$ &
1.65$_{-0.19}^{+0.16}$ & 0.85$_{-0.01}^{+0.01}$ &
1.59$_{-0.11}^{+0.12}$\\
{\tt GAUSSIAN} & $\sigma$ & keV & 0.84$_{-0.41}^{+0.13}$ & 0.65$_{-0.19}^{+0.24}$ &
0.95$_{-0.17}^{+0.17}$ & 1.07$_{-0.09}^{+0.09}$\\
{\tt GAUSSIAN} & K$_{\rm{line}}$ & phot cm$^{-2}$ s$^{-1}$ ($\times10^{-3}$) & 2.63$_{-1.68}^{+1.57}$ & 1.42$_{-0.54}^{+0.81}$ &
2.56$_{-0.65}^{+0.76}$ & 7.61$_{-0.95}^{+1.01}$\\
\noalign{\smallskip}
\hline
\noalign{\smallskip}
& $\chi^{2}$ & & 528.90 & 546.66 & 429.43 & 698.87 \\
& $\chi^{2}$/D.o.F. &  & 1.37 & 1.37 & 1.22 & 1.79 \\
\noalign{\smallskip}
\hline
\noalign{\smallskip}
\end{tabular}
\begin{list}{}{}
\item[a] The Fe-line energy E$_{\rm{line}}$ in the component {\tt GAUSSIAN} has been {\it fixed} to 6.4 keV
\item[b] Errors quoted are at the 90\% confidence level.
\end{list}
\end{table*}

The behaviour at energies above 30 keV is suggestive of the presence
of a reflected component in the spectra. To test this scenario, we
introduced the additive model {\tt PEXRIV} (Magdziarz \& Zdziarski
\cite{magdziarz}), which describes the Compton effect on a possibly
ionized disk, when the incident spectrum is well described by a power
law with a high energy cutoff. In conjunction with the {\tt PEXRIV}
component, we performed fits representing the Fe fluorescence line
with the component {\tt LAOR} (Laor \cite{laor}), which describes
the line profile from an accretion disk.

\begin{figure}
   \centering
   \includegraphics[          width=8.5cm]{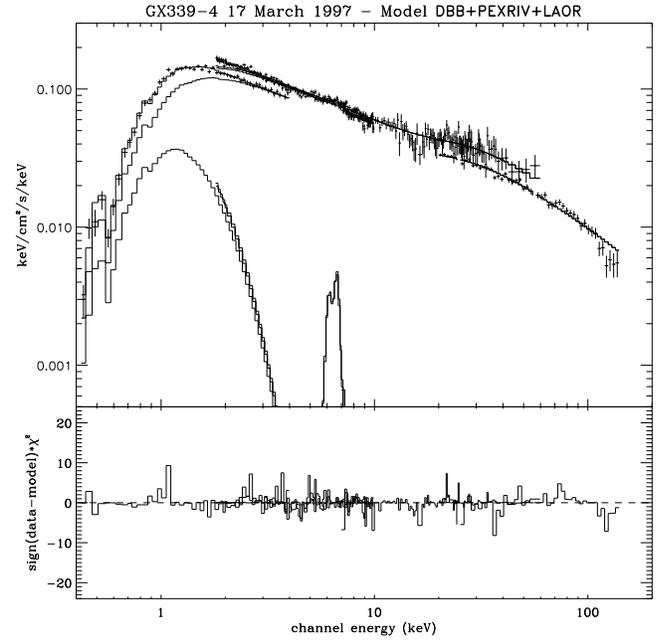}
   \caption{Result of the {\tt WABS (DISKBB + PEXRIV + LAOR)} fit to the
            spectrum of Mar 17 : unfolded spectrum in top panel,
            contribution to $\chi^2$ in bottom panel.}
    \label{FigDPL}%
    \end{figure}

Fitting results (see Table~\ref{TabDPL}) indicate clearly that the Compton reflection scenario gives
a good description of our data, and Fig.~\ref{FigDPL} shows how good is the agreement of the
assumed model with the data.

\begin{table*}
      \caption[]{Spectral fits with the {\tt WABS (DISKBB + PEXRIV + LAOR)} model}
      \label{TabDPL}
\begin{tabular}{lllllll}
\hline
\hline
\noalign{\smallskip}
\multicolumn{1}{l}{Component} &
\multicolumn{1}{l}{Parameter} &
\multicolumn{1}{l}{Unit} &
\multicolumn{1}{l}{17 March} &
\multicolumn{1}{l}{6 April} &
\multicolumn{1}{l}{6 September} &
\multicolumn{1}{l}{1 October}\\
\noalign{\smallskip}
\hline
\noalign{\smallskip}
{\tt WABS} & N$_{\rm{H}}$ & 10$^{21}$ cm$^{-2}$ & 4.76$_{-0.35}^{+0.31}$ & 4.66$_{-0.22}^{+0.14}$ &
4.46$_{-0.16}^{+0.06}$ & 5.39$_{-0.13}^{+0.14}$\\
{\tt DISKBB} & T$_{\rm{in}}$ & keV & 0.32$_{-0.04}^{+0.05}$ & 0.43$_{-0.04}^{+0.04}$ &
0.41$_{-0.03}^{+0.04}$ & 0.49$_{-0.04}^{+0.02}$\\
{\tt DISKBB} & K$_{\rm{DBB}}$ & (km/10 pc)$^{2}$ & 1847.$_{-1033}^{+2545}$ & 378.6$_{-153.5}^{+274.1}$ &
179.7$_{-40.9}^{+104.7}$ & 397.4$_{-43.3}^{+67.7}$\\   
{\tt PEXRIV} & $\Gamma$ & & 1.73$_{-0.03}^{+0.04}$ & 1.67$_{-0.03}^{+0.04}$ &
1.72$_{-0.04}^{+0.01}$ & 1.75$_{-0.01}^{+0.01}$\\
{\tt PEXRIV} & E$_{\rm{cut}}$ & keV & 533.8$_{-161.0}^{+461.8}$ & 285.5$_{-53.6}^{+101.9}$ &
802.7$_{-152.6}^{+468.5}$ & 264.8$_{-18.5}^{+21.6}$\\
{\tt PEXRIV} & Rel.Refl & $\Omega/2\pi$ & 0.80$_{-0.13}^{+0.08}$ & 0.70$_{-0.10}^{+0.12}$ &
0.88$_{-0.11}^{+0.03}$ & 0.90$_{-0.04}^{+0.04}$\\
{\tt PEXRIV} & Fe$_{\rm{abund}}$ & {\it Relative to solar abundance} & 3.68$_{-1.05}^{+0.71}$ & 4.85$_{-1.32}^{+1.51}$ &
2.85$_{-0.46}^{+0.93}$ & 3.50$_{-0.53}^{+0.47}$\\
{\tt PEXRIV} & x$_{\rm{i}}$ & & 0.37 & 4.73$\times$10$^{-2}$ & 5.55 $\times$ 10$^{-12}$ &
9.12$\times$10$^{-6}$\\
&&& $^{(<16.29)}$ & $^{(<15.31)}$ & $^{(<0.07)}$ & $^{(<36.12)}$\\
{\tt PEXRIV} & K$_{\rm{Pexriv}}$ & keV/cm$^{2}$ s keV (at 1 keV) & 0.293$_{-0.013}^{+0.014}$ & 0.237$_{-0.006}^{+0.013}$ &
0.195$_{-0.006}^{+0.003}$ & 0.416$_{-0.017}^{+0.009}$\\
{\tt LAOR} & K$_{\rm{Laor}}$ & photons cm$^{-2}$ s$^{-1}$ ($\times 10^{-4}$) & 8.33$_{-6.20}^{+3.55}$ &
5.13$_{-2.03}^{+2.84}$ & 5.33$_{-2.23}^{+2.64}$ & 16.90$_{-12.22}^{+3.24}$\\
\noalign{\smallskip}
\hline
\noalign{\smallskip}
& Flux & erg cm$^{-2}$ s$^{-1}$ ($\times 10^{-9}$) & 5.98 & 5.21 & 4.34 & 7.46\\
& Luminosity & erg s$^{-1}$ ($\times 10^{37}$) & 2.24 & 1.95 & 1.63 & 2.79\\
\noalign{\smallskip}
\hline
\noalign{\smallskip}
& $\chi^2$ & & 391.86 & 409.86 & 340.87 & 446.73\\
& $\chi^2$/D.o.F. & & 1.02 & 1.04 & 0.99 & 1.17\\
\noalign{\smallskip}
\hline
\noalign{\smallskip}
\multicolumn{7}{l}{{Cross-normalisation constants with reference to MECS3 = unity}}\\
\noalign{\smallskip}
& LECS & & 0.8145 & 0.8065 & 0.7921 & 0.8139\\
& MECS1 & & 0.9768 & 0.9758 & --- & ---\\
& MECS2 & & 0.9723 & 0.9731 & 0.9806 & 0.9743\\
& HPGSPC & & 1.056 & 1.044 & 1.048 & 1.019\\
& PDS & & 0.8116 & 0.8492 & 0.8096 & 0.7834\\
\noalign{\smallskip}
\hline
\noalign{\smallskip}
\end{tabular}
\begin{list}{}{}
\item[a] The following parameters have been fixed: Fe-line energy E$_{\rm{Laor}}$ in the component {\tt LAOR} to 
6.4 keV; the disk inclination angle, present in both components {\tt PEXRIV} and {\tt LAOR}, to the value $cos~i=$0.5.
\item[b] Errors quoted are at the 90\% confidence level.
\item[c] Reported fluxes and luminosities are relative to the 0.4--140 kev energy band. Luminosities assume a distance of 5.6 kpc.
\end{list}
\end{table*}

\section{Discussion}

Our results clearly show that the source was
in the hard-low state during all four
observations (see Table~\ref{TabDPL} for numerical results and
Fig.~\ref{FigDPL} for the unfolded spectrum).
Count rates and luminosities for each observation
are all within a factor of two from each other (see Tables
\ref{TabJou} and \ref{TabDPL}) and are typical of
GX~339-4 in its hard-low state (Zdziarski et al.
\cite{zdziarski}; Ilovaisky et al. \cite{ilovaisky};
Mendez \& van der Klis \cite{mendez}; Belloni et
al. \cite{belloni}; Wilms et al. \cite{wilms}). The
spectra are also resemblant of black hole candidates
in such spectral state, being mainly dominated by a power law, 
with a weak soft excess at lower energies
(around 1 keV). The spectral indices of such power laws
are also typical of hard-low states, and confirm that the
spectra are generated in an optically thin region,
assuming they are due to Comptonisation.
The large solid angle required by the reflection component
suggests the presence of an extended accretion disk, despite
the fact that the direct emission from it accounts only for
a small fraction of the soft flux (about 3-6\% in the 0.4-10 keV band).

Very little we can argue from the ionisation parameter:
all confidence intervals are consistent with zero,   
but at the same time have very high upper limits.
Equivalent widths of the Fe line (from Tables \ref{TabDCG}
and \ref{TabDPL}) are also consistent with previous
value in litterature (Ueda et al. \cite{ueda}; Zdziarski et al. \cite{zdziarski};
Dolan et al. \cite{dolan}), lying in the range 50--106 eV.

The values of the high energy cutoff of the incident
spectrum require some comments. Such values are
clearly outside of the energy range of BeppoSAX instruments, but their
determination is needed,
in order to constrain accurately the
reflected spectrum. They seem not
unreasonable. If the cutoff energy were inside
the BeppoSAX energy range, the
incident spectrum at higher energies would not
provide enough photons to generate such an intense
reflected component.
The absence of a cutoff below 140 keV agrees with previous
results (Dolan et al. \cite{dolan}; Bouchet et al.
\cite{bouchet}). 
In order to
confirm the presence of a high energy cutoff in the near MeV range
further investigations, e.g. with Integral, are required.

\begin{acknowledgements}
The BeppoSAX satellite has been funded and operated by the Italian Space Agency (ASI).
The Rossi XTE ASM data have been retrieved from MIT RXTE web site.
\end{acknowledgements}

\end{document}